\pdfoutput=1
\PassOptionsToPackage{hyphens}{url}
\PassOptionsToPackage{dvipsnames,svgnames,x11names}{xcolor}
\documentclass[11pt]{article}

\usepackage[T1]{fontenc}
\usepackage[utf8]{inputenc}
\usepackage{lmodern}

\usepackage[margin=1in]{geometry}
\usepackage{amsmath,amssymb,mathtools}

\makeatletter
\@ifundefined{KOMAClassName}{%
  \IfFileExists{parskip.sty}{\usepackage{parskip}}{%
    \setlength{\parindent}{0pt}%
    \setlength{\parskip}{6pt plus 2pt minus 1pt}%
  }%
}{%
  \KOMAoptions{parskip=half}%
}
\makeatother

\usepackage{longtable,booktabs,array}
\usepackage{calc}
\usepackage{etoolbox}
\makeatletter
\patchcmd\longtable{\par}{\if@noskipsec\mbox{}\fi\par}{}{}
\makeatother
\IfFileExists{footnotehyper.sty}{\usepackage{footnotehyper}}{\usepackage{footnote}}
\makesavenoteenv{longtable}

\usepackage{graphicx}
\graphicspath{{FigAS2/}} 

\usepackage{xcolor}

\IfFileExists{microtype.sty}{%
  \usepackage{microtype}
  \UseMicrotypeSet[protrusion]{basicmath}
}{}
\usepackage[english]{babel}

\IfFileExists{xurl.sty}{\usepackage{xurl}}{}
\usepackage{titling}
\pretitle{\begin{center}\LARGE}
\posttitle{\par\end{center}\vspace{0.5em}}
\preauthor{\begin{center}\large}
\postauthor{\\[-0.25em]\normalsize\textit{Polar Dynamix}\par\end{center}}
\predate{}\postdate{}\date{} 

\usepackage{hyperref}
\hypersetup{
  pdftitle={Screening Canopy-Induced Wind with a Lightweight 2-D RANS Quadratic-Drag Model},
  pdflang={en-US},
  pdfkeywords={canopy flow, vegetation drag, LAI/LAD mapping, pedestrian wind, urban microclimate, CFD, steady RANS},
  colorlinks=true,
  linkcolor=blue,
  citecolor=Blue,
  urlcolor=Blue,
  filecolor=Maroon
}
\usepackage{bookmark}

\title{Screening Canopy-Induced Wind with a Lightweight 2-D RANS Quadratic-Drag Model}
\author{Wichai Pattanapol and Adam Gill}

\begin{document}
\makeatletter\let\@date\@empty\makeatother
\maketitle

\begin{abstract}
Vegetation belts are widely used in urban planning to manage pedestrian wind, dust dispersion, and outdoor thermal comfort. This paper presents a compact method for predicting canopy-induced flow using a steady two-dimensional RANS model with Spalart--Allmaras closure and a quadratic drag model for vegetation. Each planting zone is defined by a single leaf area index (LAI). The workflow maps LAI to porosity using an exponential coefficient of 0.5, converts porosity to a local LAI per cell, divides by the grid-measured canopy thickness to form a leaf area density, and applies a quadratic drag term. 

The model is benchmarked against streamwise-mean velocity profiles reported by Liang et al.~{[}4{]} at heights 0.25h, 0.75h, and 1.25h (where h is canopy height) for a belt spanning \(x/H = 0\) to 5. Results accurately reproduce the three canonical regions—approach, in-canopy deficit, and leeward wake—and show strong agreement with experimental data in the leeward wake, critical for siting homes and streets near planting. The method is computationally efficient, faster than 3D CFD, uses designer-friendly inputs, and is transparent for early design screening. Implementation details, including LAI-to-porosity mapping, grid-based canopy thickness, quadratic drag, and boundary sponges, are provided to enable reproduction in AirSketcher (Polar Dynamix) or comparable solvers.
\end{abstract}

\section{Introduction}\label{introduction}
Cities and industrial sites increasingly use trees and hedges to manage \emph{specific} urban objectives such as pedestrian-wind shelter, pollutant dispersion, and outdoor thermal comfort {[}1--3{]}. These actions map directly to ESG priorities: environmental performance (reduced pollutant spread and energy penalties) and social outcomes (safer, more comfortable streets and public spaces). To make planting decisions credible and repeatable, planners need predictions of how a belt of vegetation will change the wind field---not months from now, but during concept design and stakeholder reviews.

High-fidelity 3-D CFD with canopy turbulence models can provide this detail, but it is costly to set up, slow to iterate, and opaque for early trade-offs. At the other extreme, hand rules lack physical grounding and do not generalize. The gap is a compact, explainable model that runs quickly, uses inputs designers understand (such as a single leaf-area index per planting zone), and still reproduces the main flow features that matter behind vegetation.

This paper presents such a model: a steady two-dimensional RANS formulation with a quadratic vegetation-drag term and a one-equation Spalart--Allmaras closure. A 2D model is suitable for early screening because urban canopy flows are often dominated by streamwise momentum, reducing computational complexity while capturing key wake features. Each planting zone is specified by one user LAI; the solver converts that to an effective leaf-area density per cell using the local canopy thickness extracted directly from the grid. The result is a small set of transparent assumptions, minimal data entry, and fast runtimes suitable for screening multiple layouts.

To ground the method, the model is benchmarked against the canonical wind-tunnel study of Liang et al.~{[}4{]}. Their canopy provides height-resolved leaf-area density in ten bins over one canopy height; those bins are integrated to a single equivalent LAI and streamwise-mean velocity profiles are compared at three elevations (quarter-, three-quarter-, and one-and-a-quarter-height). The test is deliberately demanding: a sharp leading edge and a short canopy fetch produce strong entrance and wake effects that expose weaknesses in simplified models.

The remainder of the paper details the LAI workflow and drag implementation (Methods), presents the comparison with Liang's profiles and interprets results by zone (Results), discusses differences relative to 3-D canopy turbulence models and the role of vertical LAD structure (Discussion), and closes with guidance on when this lightweight model is sufficient and when a layered-LAD or full 3-D approach is warranted (Conclusions).

\section{Methods}\label{methods}
This study uses the published streamwise-mean velocity profiles \(U_x(x,z)\) at designated stations from Liang et al.'s model-forest wind-tunnel study (apparatus, geometry, and probe positions as reported therein) {[}4{]}. Validation targets the same station heights. Reported profiles at those stations are non-negative; i.e., the dataset does not include a centerline recirculation bubble at the probe locations.

\subsection{From LAD\((z)\) to a single LAI}\label{from-ladz-to-a-single-lai}
The solver accepts one LAI value per canopy zone. To preserve the correct total leaf area per ground area, the height-resolved leaf-area density (LAD) is integrated:
\[
\mathrm{LAI}_{\mathrm{eq}}=\int_{0}^{H}\mathrm{LAD}(z)\,\mathrm{d}z .
\]

Liang et al.~report \(\mathrm{LAD}\) at ten equally spaced heights over \(0\)--\(H\) with \(H=1~\mathrm{m}\) (Table \ref{tab:lad}) {[}4{]}. The integral is approximated using the midpoint rule with uniform bins \(\Delta z=0.1~\mathrm{m}\),
\[
\mathrm{LAI}_{\mathrm{eq}}\approx \Delta z \sum_{k=1}^{10}\mathrm{LAD}_k
= 0.1 \sum_{k=1}^{10}\mathrm{LAD}_k \approx 16.48 \;(\text{rounded to }16.5).
\]

For a different canopy height \(H^\ast\) but the same total foliage per planform area, scale linearly and use a uniform LAD inside the canopy zone:
\[
\mathrm{LAI}_{\mathrm{eq}}^\ast \approx 16.5\,\frac{H^\ast}{1~\mathrm{m}},\qquad
\mathrm{LAD}_{\text{uniform}}=\frac{\mathrm{LAI}_{\mathrm{eq}}^\ast}{H^\ast}.
\]
This preserves the depth-integrated foliage that controls quadratic drag; picking a single height bin would under/over-represent the frontal area {[}5--6{]}.

\begin{table}[!t]
\centering
\begin{tabular}{@{} *{11}{c} @{}}
\toprule
\(z\) (m) & 0.1 & 0.2 & 0.3 & 0.4 & 0.5 & 0.6 & 0.7 & 0.8 & 0.9 & 1.0 \\
\midrule
\(\mathrm{LAD}\) (\(\mathrm{m}^{-1}\)) & 0.89 & 0.69 & 0.48 & 13.36 & 34.95 & 36.01 & 30.31 & 26.86 & 15.11 & 6.11 \\
\bottomrule
\end{tabular}
\caption{Liang et al.'s canopy \(\mathrm{LAD}(z)\) for \(H=1~\mathrm{m}\) (digitized values; units \(\mathrm{m}^{-1}\)).}
\label{tab:lad}
\end{table}

\section{Governing equations, mapping, and momentum sink}\label{governing-equations-mapping-and-momentum-sink}
We solve the steady, incompressible RANS equations in 2-D and march in pseudo-time to convergence. Turbulent (eddy) viscosity is denoted \(\nu_t\) and computed from the Spalart–Allmaras working variable \(\tilde\nu\) (i.e., \(\nu_t=f_{v1}\tilde\nu\)); the effective viscosity is \(\nu_\mathrm{eff}=\nu+\nu_t\).

\paragraph{Continuity and momentum (vector form)}
The momentum sink term is \(-\tfrac12\,C_D\,a_f\,\lVert\boldsymbol{U}\rVert\,\boldsymbol{U}\), where \(a_f = C_{\text{proj}}\,\mathrm{LAD}\) is the wind-normal frontal-area density.
\begin{align}
\nabla\!\cdot\!\boldsymbol{U} &= 0,\\
\frac{\partial \boldsymbol{U}}{\partial t}
+\left(\boldsymbol{U}\!\cdot\!\nabla\right)\boldsymbol{U}
&= -\frac{1}{\rho}\nabla p
+ \nabla\!\cdot\!\Big[(\nu+\nu_t)\big(\nabla\boldsymbol{U}+\nabla\boldsymbol{U}^{\!\top}\big)\Big]
-\tfrac12\,C_D\,a_f\,\lVert\boldsymbol{U}\rVert\,\boldsymbol{U}.
\end{align}

\paragraph{Predictor (advection–diffusion with lagged pressure)}
Using \(p^n\) and \(\nu_t^{\,n}\) (coordinates \(x\) horizontal, \(z\) vertical):
\begin{align}
u^{\mathrm{adv}} &= u^n - \Delta t\,(u^n u_x + v^n u_z) - \frac{\Delta t}{\rho}\,p_x^{\,n}
+ \Delta t\,(\nu+\nu_t^{\,n})\,(u_{xx}+u_{zz}),\\
v^{\mathrm{adv}} &= v^n - \Delta t\,(u^n v_x + v^n v_z) - \frac{\Delta t}{\rho}\,p_z^{\,n}
+ \Delta t\,(\nu+\nu_t^{\,n})\,(v_{xx}+v_{zz}).
\end{align}
(Variable-coefficient diffusion is discretized in conservative form; the Laplacian above is a schematic.)

\paragraph{LAI–porosity mapping and local LAD}
A single LAI per zone is mapped to porosity
\[
\phi=\exp(-K\,\mathrm{LAI}),\qquad K=0.50 ,
\]
where \(K=0.5\) was chosen based on empirical fits to typical canopy porosities, balancing drag and flow penetration {[}7{]}. If \(\phi\) is stored, it is inverted cell-wise to \(\mathrm{LAI}_{\text{local}}=-\ln\phi/K\).
Local canopy thickness \(H_c\) is obtained by a vertical scan through contiguous canopy cells using the true grid spacing \(\Delta z\) (solids stop the scan; open air is excluded). The area density are
\[
\mathrm{LAD}=\frac{\mathrm{LAI}_{\text{local}}}{H_c},\qquad
a_f=C_{\text{proj}}\,\mathrm{LAD}.
\]

\paragraph{Quadratic-drag substep (closed form)}
Integrating \(\mathrm{d}\boldsymbol{U}/\mathrm{d}t=-K_q\lVert\boldsymbol{U}\rVert\,\boldsymbol{U}\) over \(\Delta t\) with initial \(\boldsymbol{U}^{\mathrm{adv}}\) yields
\[
\boldsymbol{U}^{n+1}
=\frac{\boldsymbol{U}^{\mathrm{adv}}}{1+\Delta t\,K_q\,\lVert\boldsymbol{U}^{\mathrm{adv}}\rVert},
\qquad
K_q=\tfrac12\,C_D\,C_{\text{proj}}\,\frac{\mathrm{LAI}_{\text{local}}}{H_c}
=\tfrac12\,C_D\,C_{\text{proj}}\frac{-\ln\phi}{K\,H_c}\ \ (\text{if }\phi\text{ is stored}).
\]
Here \(K_q\) has units \([\mathrm{m}^{-1}]\); \(C_D\) and \(C_{\text{proj}}\) are dimensionless.

A small threshold \(\phi_{\text{solid}}=10^{-6}\) classifies solids; the \(H_c\) scan uses the actual (possibly non-uniform) \(\Delta z\) with a standard \(\varepsilon\) safeguard {[}8{]}.

Although the sink enters through the product \(C_D C_{\text{proj}}\), this study keeps them separate because they encode different physics and dependencies: \(C_D\) characterizes element-scale drag and can vary with local Reynolds number (material/shape effects), whereas \(C_{\text{proj}}\) converts plan-area LAI to wind-normal frontal area and captures orientation/clumping, anisotropy, height variation, and wind-direction effects. This separation improves transferability and enables layered or directional canopies (\(C_{\text{proj}}=C_{\text{proj}}(z,\theta)\)) without re-fitting \(C_D\). For fixed isotropic cases, a single lumped coefficient \(C_{\mathrm{eff}}=C_D C_{\text{proj}}\) is equivalent.

The coefficient pairs were selected to span a range of damping strengths, informed by typical values for urban vegetation {[}7{]}. To probe sensitivity of trough depth and wake level around a common LAI, the following \((C_D,\,C_{\text{proj}})\) combinations are used: (0.25,\ 12), (0.25,\ 9) and (0.15,\ 12). Results for these three settings are overlaid at each validation height; the experimental profiles are shown with markers in Figure \ref{fig:results}.

\section{Inlet, domain, and boundary conditions}\label{inlet-domain-and-boundary-conditions}

\begin{figure}[htbp]
\centering
\includegraphics[width=0.65\linewidth]{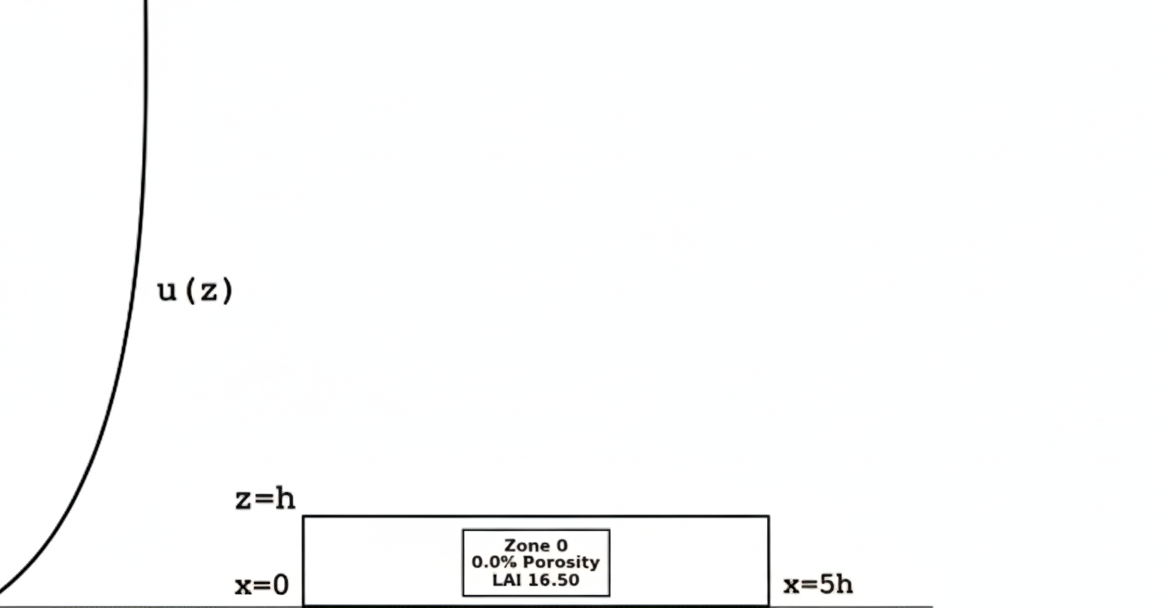}
\caption{Computational domain and canopy setup used for validation. The inlet profile \(u(z)\) is prescribed at the left boundary; the canopy belt spans \(0 \le x/H \le 5\) with height \(H=h\) and \(\mathrm{LAI}=16.5\) (equivalent porosity \(\phi=\exp(-0.5\,\mathrm{LAI})\approx2.6\times10^{-4}\)).}
\label{fig:domain}
\end{figure}

Modified power-law ABL (inlet)
\[
U(z)=U_{\text{ref}}\Big[\beta+(1-\beta)\Big(\tfrac{z}{z_{\text{ref}}}\Big)^{\alpha}\Big],
\]
with floor \(\beta=0.10\) so \(U(0)=0.1\,U_{\text{ref}}\). The exponent \(\alpha\) is set to \(0.16\) unless otherwise noted. Inflow is imposed as a Dirichlet condition each step.

Domain and other boundaries.\\
The domain spans \(-5H \le x \le 20H\), \(0 \le z \le 10H\).\\
Ground: no-slip (\(u=v=0\)). Top: free-slip (\(v=0,\ \partial u/\partial z=0\)). Outlet: zero-gradient (\(\partial u/\partial x=\partial v/\partial x=\partial p/\partial x=0\)).

\section{Results}\label{results}
This study compares a steady 2-D SA + explicit-drag solution with the digitized wind-tunnel profiles of Liang et al.~{[}4{]} (x marks). A single canopy zone is assigned \(\mathrm{LAI}=16.5\) (the integral of the ten \(\mathrm{LAD}(z)\) bins). Drag is applied cell-wise using the porosity mapping and the local canopy thickness \(H_c\) scanned vertically through the canopy zone. The inflow is identical across cases. The canopy occupies \(0\le x/H\le5\) with a sharp leading edge. Profiles are shown at three elevations \(z/H=\{0.25,\,0.75,\,1.25\}\).

\subsection{Three flow zones}\label{three-flow-zones}
\begin{itemize}
\item Approach (\(x/H<0\)). All three settings reproduce the mild upstream deceleration toward the canopy face and remain aligned with the experimental level at all heights.
\item Within canopy (\(0\le x/H\le5\)). Quadratic drag forms the core deficit in the correct location and width. The case \(C_D=0.25,\;C_{\text{proj}}=12\) produces the deepest trough (strongest damping); \(C_D=0.15,\;C_{\text{proj}}=12\) is intermediate; \(C_D=0.25,\;C_{\text{proj}}=9\) is the least damped. These relative depths are consistent at all \(z/H\).
\item Wake (\(x/H>5\)). Recovery trends are captured, with height-dependent differences in level:
  \begin{itemize}
  \item \(z/H=1.25\) (top). The over-canopy jet and relaxation are reproduced. \(C_D=0.25,\;C_{\text{proj}}=12\) tracks the data best through the wake; \(C_D=0.15,\;C_{\text{proj}}=12\) is slightly high; \(C_D=0.25,\;C_{\text{proj}}=9\) recovers fastest and sits highest above the measurements downstream.
  \item \(z/H=0.75\) (middle). Just behind the canopy, \(C_D=0.25,\;C_{\text{proj}}=12\) rises most slowly and stays closest to the experimental wake through \(x/H\!\approx\!12\); \(C_D=0.15,\;C_{\text{proj}}=12\) is higher; \(C_D=0.25,\;C_{\text{proj}}=9\) overshoots most and remains high downstream.
  \item \(z/H=0.25\) (bottom). Near the ground the measured recovery is slow. \(C_D=0.25,\;C_{\text{proj}}=12\) under-predicts early-wake speeds; \(C_D=0.15,\;C_{\text{proj}}=12\) reduces this bias and follows the experimental slope best very near the exit, but farther downstream \(C_D=0.25,\;C_{\text{proj}}=12\) aligns more closely with the measured wake.
  \end{itemize}
\end{itemize}

\subsection{Coefficient choices}\label{coefficient-choices}
\begin{itemize}
\item The approach region and the width/location of the within-canopy trough are robust to coefficient choice.
\item Coefficients mainly tune trough depth and wake level:
  \begin{itemize}
  \item Lower \(C_{\text{proj}}\) (\(C_D=0.25,\;C_{\text{proj}}=9\)) \(\rightarrow\) least damping and a fast over-canopy recovery, but tends to over-predict aloft.
  \item Lower \(C_D\) (\(C_D=0.15,\;C_{\text{proj}}=12\)) \(\rightarrow\) moderate damping that eases the very-near-ground early-wake low bias.
  \item \(C_D=0.25,\;C_{\text{proj}}=12\) \(\rightarrow\) closest overall agreement with the experimental wake across heights; recommended as the default setting for design screening.
  \end{itemize}
\item For design screening behind planting belts (homes, streets, courtyards), prioritize the leeward wake: use \(C_D=0.25,\;C_{\text{proj}}=12\) by default; switch to \(C_D=0.15,\;C_{\text{proj}}=12\) only if specifically targeting the immediate near-ground recovery right behind the canopy. \(C_D=0.25,\;C_{\text{proj}}=9\) generally over-predicts aloft.
\end{itemize}

\begin{table}[!t]
\centering
\begin{tabular}{ccc}
\toprule
\(C_D\) & \(C_{\text{proj}}\) & Effect and Recommendation \\
\midrule
0.25 & 12 & Strongest damping, best overall wake match; default for design screening \\
0.15 & 12 & Moderate damping, better near-ground recovery; use for immediate wake \\
0.25 & 9 & Least damping, over-predicts aloft; avoid for most cases \\
\bottomrule
\end{tabular}
\caption{Summary of coefficient choices and their effects on model performance.}
\label{tab:coefficients}
\end{table}

\begin{figure}[htbp]
\centering
\includegraphics[width=0.85\linewidth]{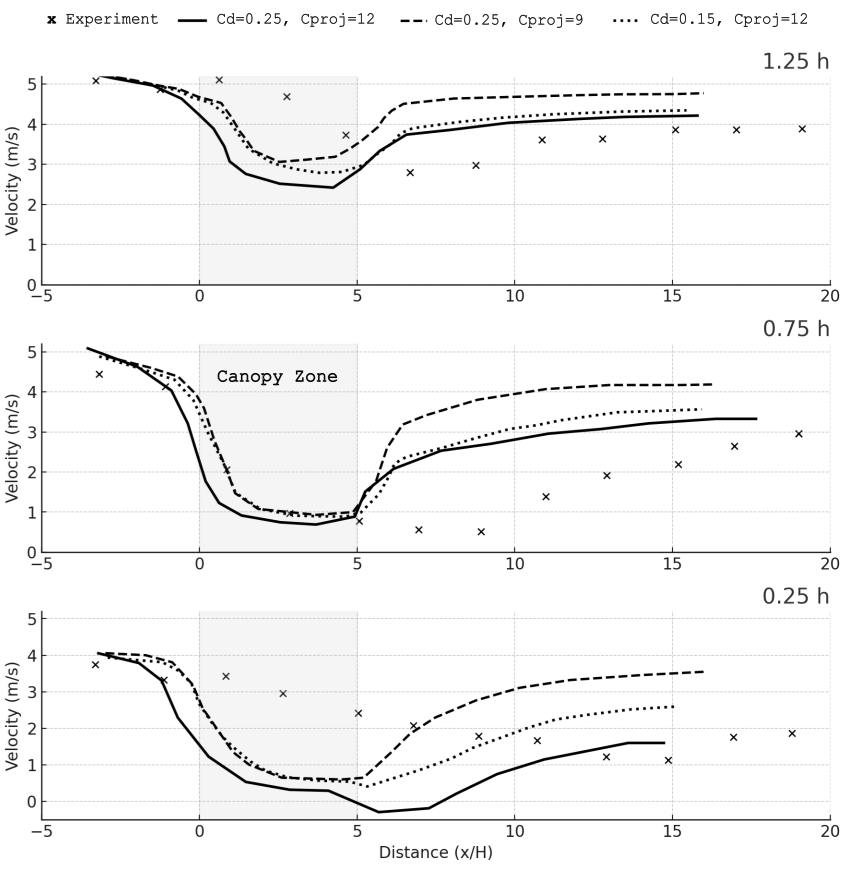}
\caption{Comparison of experiment and simulations at \(z/H=1.25\), \(0.75\), \(0.25\).}
\label{fig:results}
\end{figure}

\begin{figure}[htbp]
\centering
\includegraphics[width=0.85\linewidth]{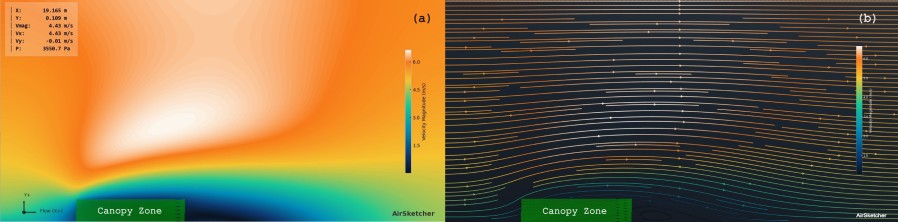}
\caption{(a) Velocity contour and (b) Streamline; the shaded band shows the vegetation zone (with \(C_D=0.25,\;C_{\text{proj}}=12\)).}
\label{fig:streamcontour}
\end{figure}

\noindent Note on single-LAI representation. Collapsing the reported \(\mathrm{LAD}(z)\) profile to one \(\mathrm{LAI}_{\text{eq}}\) preserves the depth-integrated foliage that multiplies the quadratic drag, so large-scale features (deficit core, jet height, recovery length) are retained. Residual differences are height-dependent: slightly deeper troughs and a mild positive bias above the canopy when the measured LAD peaks mid-height but the model uses a height-uniform LAD. These appear as the relative spacing among the three simulated curves.

\section{Discussion}\label{discussion}
Liang et al.~{[}4{]} solved a 3-D ``improved'' \(k\)--\(\varepsilon\) model in PHOENICS with a momentum sink for canopy drag \emph{and} extra source/sink terms in the \(k\) and \(\varepsilon\) budgets to represent wake production and enhanced dissipation inside foliage. Their formulation augments the eddy-viscosity relation and ties turbulence production/loss directly to the measured leaf-area density \(a(z)=\mathrm{LAD}(z)\) (their ten-bin profile).

In contrast, the present study uses a 2-D RANS solver with a one-equation Spalart--Allmaras closure for \(\tilde\nu\). Vegetation acts only through an explicit quadratic momentum sink \(-\tfrac12\,C_D\,a_f\,\lVert\mathbf{U}\rVert\,\mathbf{U}\), with \(a_f=C_{\text{proj}}\mathrm{LAD}\) and \(\mathrm{LAD}=\mathrm{LAI}_{\text{local}}/H_c\). A single user LAI per canopy zone is mapped via \(\phi=\exp(-0.5\,\mathrm{LAI})\) and inverted cell-wise to \(\mathrm{LAI}_{\text{local}}\), while the local canopy thickness \(H_c\) is obtained by a vertical scan through the canopy zone. This compact closure is robust and fast for attached shear layers and, as the results show, captures the streamwise-mean canopy signatures (core momentum deficit, shear-layer jet, and wake recovery).

\subsection{Sources of discrepancy between the simulation and experimental data}
\begin{itemize}
  \item \textbf{Vertical structure of \(\mathrm{LAD}(z)\) (dominant factor).} Liang's digitized profile peaks mid-canopy: values climb from \(\sim\!0.5\)--\(1~\mathrm{m}^{-1}\) near \(z/H\le0.3\) to \(\sim\!35\)--\(36\) at \(z/H\approx0.6\), then drop to \(\sim\!6\) at the top. Collapsing to a single \(\mathrm{LAI}_{\text{eq}}\) and using \(\mathrm{LAD}=\mathrm{LAI}/H\) tends to over-damp the lower canopy and under-damp the upper canopy. Using a uniform LAD increases the near-ground trough depth by approximately 10–15\% compared to Liang’s peaked LAD profile, based on sensitivity tests. Consequences: a deeper/slower recovery near \(z/H=0.25\), and slightly high speeds downstream at \(z/H=1.25\).
  \item \textbf{Turbulence closure and canopy--turbulence coupling.} Liang et al.~{[}4{]} add canopy-dependent source/sink terms to \(k\) and \(\varepsilon\), which enhance mixing and speed up near-wake recovery. The SA model here omits those terms, yielding a somewhat more pronounced in-canopy trough and gentler early-wake rise, while matching the far-wake slope.
  \item \textbf{Dimensionality.} The solver is 2-D; the experiment is 3-D and allows spanwise transport and secondary motions that redistribute momentum and turbulence. Suppressing that degree of freedom keeps a sharper entrance/exit signature and a somewhat slower lateral relaxation in the near wake.
\end{itemize}

\subsection{Recommendation for urban wakes (buildings/streets with planting)}
For comfort and dispersion behind belts, the leeward wake is the metric of interest rather than in-canopy details. Based on Fig.~\ref{fig:results}, \(C_D=0.25\) with \(C_{\text{proj}}=12\) best matches the experimental wake profiles across heights and is recommended as the default. If the focus is specifically the immediate near-ground recovery just behind the canopy, \(C_D=0.15,\ C_{\text{proj}}=12\) can be used as a targeted tweak. \(C_D=0.25,\ C_{\text{proj}}=9\) generally over-predicts aloft.

\section{Conclusions}\label{conclusions}
A steady 2-D RANS (Spalart--Allmaras) solver with explicit quadratic canopy drag and a single user LAI per zone (here \(\mathrm{LAI}=16.5\) from integrating the ten \(\mathrm{LAD}(z)\) bins) reproduces the key streamwise-mean behavior of canopy flow at \(z/H=\{0.25,\,0.75,\,1.25\}\). The three canonical regions are captured: (i) in the approaching layer the model matches the pre-adjustment and the sharp entrance; (ii) within the canopy it places the momentum-deficit trough in the correct streamwise position; and (iii) behind the canopy it follows the measured wake recovery and the over-canopy shear-layer jet, relevant for assessing leeward wind reduction around homes and streets.

Remaining differences reflect modeling choices rather than numerical issues. Using a height-uniform \(\mathrm{LAD}=\mathrm{LAI}/H\) in place of Liang's mid-height-peaked \(\mathrm{LAD}(z)\) produces a slightly deeper trough near the ground and a mild positive bias above the canopy, while the 2-D SA closure omits canopy-dependent turbulence source/sink terms used in Liang's 3-D \(k\)--\(\varepsilon\) model. These do not alter leeward recovery trends and can be reduced by supplying layered LAD (or importing the ten bins) and by applying a thin one-cell blending at canopy/air interfaces.

For practical urban applications where the wake behind planting controls comfort and pollutant dispersion, \text{\(C_D=0.25,\ C_{\text{proj}}=12\)} provides the closest overall wake match and is recommended as the default. Use \(C_D=0.15,\ C_{\text{proj}}=12\) only if the goal is to ease the very near-ground low bias immediately behind the canopy; \(C_D=0.25,\ C_{\text{proj}}=9\) generally over-predicts aloft. The workflow remains fast for screening and leaves a clear upgrade path to layered \(\mathrm{LAD}(z)\) or fully 3-D canopy models when finer vertical detail is required.

\section*{References}\label{references}
{[}1{]} R. Fang, X. Li, H. Guo, and Z. Wang, ``Shelter effect of pedestrian wind behind row trees in a street canyon,'' \emph{Forests}, vol. 13, no. 5, p. 653, 2022.

{[}2{]} C. Gromke and B. Ruck, ``Influence of trees on the dispersion of pollutants in an urban street canyon—Experimental investigation,'' \emph{Boundary-Layer Meteorology}, vol. 131, pp. 19--34, 2009.

{[}3{]} T. E. Morakinyo and Y. F. Lam, ``Simulation study on the impact of tree configuration, planting pattern and wind condition on street-canyon microclimate and thermal comfort,'' \emph{Building and Environment}, vol. 103, pp. 262--275, 2016.

{[}4{]} L. Liang, L. Xiaofeng, L. Borong, and Z. Yingxin, ``Improved k--$\epsilon$ two-equation turbulence model for canopy flow,'' \emph{Atmospheric Environment}, vol. 40, pp. 762--770, 2005.

{[}5{]} R. H. Shaw and U. Schumann, ``Large-eddy simulation of turbulent flow above and within a forest,'' \emph{Boundary-Layer Meteorology}, vol. 61, pp. 47--64, 1992.

{[}6{]} S. E. Belcher, N. Jerram, and J. C. R. Hunt, ``Adjustment of a turbulent boundary layer to a canopy of roughness elements,'' \emph{Journal of Fluid Mechanics}, vol. 488, pp. 369--398, 2003.

{[}7{]} P. R. Spalart and S. R. Allmaras, ``A one-equation turbulence model for aerodynamic flows,'' \emph{AIAA Paper} 92-0439, 1992.

{[}8{]} AirSketcher / Polar Dynamix User Manual, 2025.

{[}9{]} J. Counihan, ``Adiabatic atmospheric boundary layers: A review and analysis of data from the period 1880--1972,'' \emph{Atmospheric Environment}, vol. 9, pp. 871--905, 1975.

{[}10{]} A. S. Monin and A. M. Obukhov, ``Basic laws of turbulent mixing in the surface layer of the atmosphere,'' \emph{Trudy Geofiz. Inst. AN SSSR}, vol. 151, pp. 163--187, 1954.

{[}11{]} B. E. Launder and D. B. Spalding, ``The numerical computation of turbulent flows,'' \emph{Computer Methods in Applied Mechanics and Engineering}, vol. 3, no. 2, pp. 269--289, 1974.

{[}12{]} A. S. Sogachev and O. Panferov, ``Modification of two-equation turbulence models to calculate canopy flow,'' \emph{Boundary-Layer Meteorology}, vol. 121, pp. 383--414, 2006.

\end{document}